\documentclass[12pt,preprint]{aastex}

\usepackage{graphicx}
\usepackage{natbib}
\usepackage{latexsym}

\bibpunct{(}{)}{;}{a}{}{,}

\begin{document} 

\title{A Survey of Metal Lines at High-redshift (II) : SDSS Absorption Line Studies - OVI line density, space density and gas metallicity at z$_{abs} \sim$3.0}
\author{S.~Frank\altaffilmark{1},
        S.~Mathur\altaffilmark{1},
        M. Pieri\altaffilmark{1},
        \and D.~G.~York\altaffilmark{2}}

\altaffiltext{1}{\small Department of Astronomy, Ohio State University, 140 W.18th Ave., Columbus, OH 43210, USA}
\altaffiltext{2}{\small Department of Astronomy and Astrophysics, University of Chicago, 5640 S.Ellis Avenue, Chicago, Illinois 60637, USA}

\email{frank@astronomy.ohio-state.edu}

%\date{Received ??? / Accepted ???}

\begin{abstract} 
We have analysed a large data set of \ion{O}{6}{} absorber candidates found in the spectra of 3702 SDSS quasars, focusing on a subsample of 387 AGN sightlines with an average S/N$\geq 5.0$, allowing for the detection of absorbers above a rest-frame equivalent width limit of W$_{r} \geq 0.19 \AA${} for the \ion{O}{6}{} 1032 \AA{} component.\\
Accounting for random interlopers mimicking an \ion{O}{6}{} doublet, we derive for the first time a secure lower limit for the redshift number density $\Delta N / \Delta z$ for redshifts $z_{abs} \geq 2.8$. With extensive Monte Carlo simulations we quantify the losses of absorbers due to blending with the ubiquitous Ly$\alpha${} forest lines, and estimate the success rate of retrieving each individual candidate as a function of its redshift, the emission redshift of the quasar, the strength of the absorber and the measured S/N of the spectrum by modelling typical Ly forest spectra. These correction factors allow us to derive the 'incompleteness and S/N corrected' redshift number densities of \ion{O}{6} absorbers :$\Delta N _{OVI, c} / \Delta z_{c} (2.8 < z < 3.2) = 4.6 \pm 0.3$, $\Delta N _{OVI, c} / \Delta z_{c} (3.2 < z < 3.6) = 6.7 \pm 0.8$,and $\Delta N _{OVI, c} / \Delta z_{c} (3.6 < z < 4.0) = 8.4 \pm 2.9$.\\
We can place a secure lower limit for the contribution of \ion{O}{6}{} to the closure mass density at the redshifts probed here: $\Omega _{OVI} (2.8 < z < 3.2) \geq 1.9 \times 10^{-8} h^{-1}$. We show that the strong lines we probe account for over 65\% of the mass in the \ion{O}{6}{} absorbers; the weak absorbers, while dominant in line number density, do not contribute significantly to the mass density.\\
Making a conservative assumption about the ionisation fraction, $\frac{OVI}{O}${}, and adopting the \citet[]{anders1989}{} solar abundance values, we derive the mean metallicty of the gas probed in our search : $\zeta (2.8 < z < 3.2) \geq 3.6 \times 10^{-4} h$, in good agreement with other studies. These results demonstrate that large spectroscopic datasets such as SDSS can play an important role in QSO absorption line studies, in spite of the relatively low resolution.
\end{abstract}

% \titlerunning{The running title} \authorrunning{S. Frank et
%%al.}  
%%\maketitle
%
\section{Introduction}\label{introduction}
The cosmic evolution of the majority of the baryons, as probed by absorption features of neutral hydrogen towards the sightlines of QSOs, has come to be well understood over the past years of extensive research : the availability of powerful echelle spectrographs on 8m-class telescopes like HIRES/Keck and UVES/VLT has opened the window for high resolution studies of the Lyman $\alpha${} forest down to very low column densities \citep[]{hu1995, lu1996, kim1997, kirkman1997}. On the theoretical front, models incorporating gas dynamics, radiative cooling, and photoionisation have been spectacularly successful in reproducing the observed properties of the parts of the gas traced by neutral HI throughout much of the history of the universe \citep[]{cen1994, miraldaescude1996, hernquist1996, petitjean1995, dave1997}. The emerging picture paints baryonic gas being encountered in a wide variety of physical conditions. Most of the high redshift Lyman $\alpha${} forest, which is the main repository for baryons at that time of the universe, exhibits relatively cool temperatures (T$\sim 10^{4}${} K) in low-to-medium overdensity structures, governed by photoionisation, but there are clear signs of a additional highly ionised component, traced by the prominent absorption lines of CIV and OVI.\\
The existence of metal absorption lines in the spectra of quasars, already well established early on in the study of QSO absorption lines \citep[]{bahcall1968}, and then later linked to the forest (\ion{C}{4}{} absorbers seen at the same redshift as Lyman $\alpha${} absorbers by \citet[]{meyer1987}, poses interesting problems, and in fact allows to address several astrophysical phenomena in a unique setting. First of all, metals allow insight into the structure of the intergalactic medium (IGM) itself : its temperature and small-scale velocity structures can be probed via the width of such lines. Secondly, the existence of metal absorbers reveals the impact of star formation on the IGM - either the metals had to be produced $\it in{} situ${} by Population III stars in small-scale regions, or a transport mechanism from galaxies like galactic winds into the IGM has to be invoked. Furthermore, studying the ratios of metal column densities of different species and ionic stages can be used to constrain the spectral shape of the metagalactic UV background.\\
The hot phase of the IGM (WHIM) with temperatures above T $\sim 10^{5}${} K is still comparatively poorly understood. Because collisional ionisation of hydrogen becomes significant at these temperatures, Lyman $\alpha${} becomes a worse tracer of this phase, and thus budgets of the IGM content based solely on hydrogen will underestimate the contribution of the hot gas \citep[]{simcoe2004}. The high ionisation potential and high abundance renders oxygen a prime candidate for studying the IGM phase under these conditions, possibly tracing shock-heated material. Furthermore, OVI has been long recognised as ideal candidate to trace the low density component (log N$_{HI} < 14.0$) of the photoionised Lyman forest at z$>2${} with optical ground-based spectroscopy \citep[]{hellsten1998}, with \ion{C}{4}{} being another possible tracer. The low density in this regime allows for the production of OVI via photoionisation by the intergalactic UV and soft X-ray radiation field \citep[]{haardt1996}.\\
Searches for the signatures of intervening OVI absorbers in QSO spectra have focused on two techniques. A direct line detection, based upon the identification of the characteristic OVI doublet at rest-frame wavelengths of 1032 and 1038 \AA, at low redshifts must be performed via space-based observations, and ground-based detections hence become possible only for absorber redshifts beyond $z_{abs} \sim 2.0$, when the lines are shifted redward of the atmospheric cutoff at 3000 \AA.  \citet[]{burles1996} conducted the first systematic survey for OVI absorbers at $z_{abs} \sim$1.0 with FOS aboard HST, studying a sample of 11 QSOs and identifying 9 good candidates, thus establishing that the number density per unit redshift of these absorbers is similar to, if not greater than, that for CIV and MgII absorbers at the same redshift. More recently, \citet[]{thom2008_I, thom2008_II}{} have used HST/STIS to search for intervening OVI absorbers towards 16 QSOs at $z_{em} = 0.17 - 0.57$, retrieving a sample of 27 candidates with rest-frame equivalent widths $W_{r} \geq$5m\AA. The biggest sample of such absorbers at low $z_{abs}${} has been collected by \citet[]{danforth2007}, who obtain 83 OVI systems by analysing 650 Ly$\alpha${} absorbers with HST and FUSE spectroscopy, enabling sound statistical analyses of the population.\\
Towards higher redshift, despite the advantage of having the features shifted into the wavebands for ground-based optical spectroscopy, this search method is hampered by the ever increasing density of the Ly$\alpha${} forest, rendering an unambiguous detection of the absorber doublet in its midst more difficult. Hence, most searches have employed high-resolution and high signal-to-noise (S/N) spectroscopy of a handful of QSO sightlines, to build up samples of such OVI absorbers. In order to ensure the highest likelihood of detecting the doublet securely, these studies have focused on absorber and emitter redshifts below $z \sim$3.0. To date, the number of securely identified OVI absorber components with this method in the redshift range $2.0 \leq z_{abs} \leq 3.0${} is of order 350 \citep[]{simcoe2004, simcoe2006, bergeron2005, carswell2002, fox2007, lopez2007, fox2008}.\\
 An alternative approach to characterising the metal content of the Lyman forest is the pixel-to-pixel-correlation method, pioneered by \citet[]{cowie1998}{} for CIV. With this method, statistical correlations between the mean optical depths of various transitions enable an estimate of the metallicity of low density gas. Over the same redshift range as mentioned for the direct line identification searches, \citet[]{songaila1998, schaye2000} and most recently \citet[]{aracil2005}{} assert the existence of OVI in gas of optical depth as low as $\tau _{HI} = 0.1$. For a summary of the results and a discussion of the ambiguity in their interpretations see \citet[]{aguirre2002} and \citet[]{pieri2004}.\\
In the first paper of our series \citep[]{frank2007}, we have described how we utilise the large number of high-redshift QSOs in the Sloan Digital Sky Survey (SDSS) spectroscopic sample to explore a different locus of the parameter space for direct searches : low resolution ($R \sim 1800$) and low to medium S/N at emission redshifts $z_{em} \geq 2.8$. Frank et al. (2010) describes the sample selection, the search algorithm and presents the candidates found in over 3700 sightlines. While it is clear that the lower quality of the spectra as compared to the high-resolution studies does not allow us to probe such weak absorbers as in the above mentioned searches, the large combined pathlength covered by the SDSS QSOs' sightlines enables us to systematically build up a large sample of strong \ion{O}{6}{} absorbers at unprecedented high redshifts.\\
In this paper, we explore the statistics of the subsample of the candidates towards sightlines with the best S/N, and derive, for the first time at redshift beyond $z_{abs} \geq 2.8$, secure lower limits for the redshift path density, dn/dz, of strong \ion{O}{6}{} absorbers. Furthermore, by a Monte-Carlo analysis we can determine the fraction of absorbers lost due to noise and the blending with ubiquitious Lyman forest lines, and hence derive a good estimate for the path density probed with our analysis (section 2). Enforcing a strict lower limit to the {\it total}{} column density found for our sample, we derive in section 3 a limit for the total mass density of \ion{O}{6}{} at these high redshifts, and then in section 4 we estimate the minimum metallicity of the gas traced by our absorber sample by limiting the ionisation of oxygen to the maximum value achieved in both the collisionally and photoionised cases. We compare our results with other studies, and conclude in section 5, where we also propose different scenarios for the origin of these absorbers and point out the need for higher resolution and/or signal-to-noise spectra in order to discriminate between them.\\
Unless specifically noted otherwise, we use a cosmology with H$_{0}$=71 km s$^{-1}$Mpc$^{-1}$, $\Omega _{M}$ = 0.27 and $\Omega _{\lambda}$=0.73 in this study. Abundances are given by number relative to hydrogen, and we utilise the \citet[]{anders1989}{} solar abundance values. 

%%%%%%%%%%%%%%%%%%%%%%%%%%%%%%%%%%%%%%%%%%%%%%%%%%%%
%%%%%%%%%%%%%%%%%%%%%%%%%%%%%%%%%%%%%%%%%%%%%%%%%%%%

\section{Frequency of \ion{O}{6} absorbers}
\subsection{The uncorrected Redshift Number Density}

%%%%%%%%%%%%%%%%%%%%%%%%%%%%%%%%%%%%%%%%%%%%%%%%%%%%
%%%%%%%%%%%%%%%%%%%%%%%%%%%%%%%%%%%%%%%%%%%%%%%%%%%%

The redshift number density of \ion{O}{6}{} absorbers dn/dz, defined as the number of absorbers per unit redshift, is a function of the limiting equivalent width $W_{r, limit}$. Here we estimate dn/dz as :
\begin{equation}
dn/dz (W_{r, limit}) = \frac{n_{tot} (>W_{r, limit}) - n_{rand} (>W_{r, limit})}{\Delta z _{tot}} 
\end{equation} 
where $n_{tot} (>W_{r, limit})$ is the total number of absorbers found by the search procedure described in \citet[]{frank2007}, $n_{rand} (>W_{r, limit})$ is the estimated number of random interlopers and false classifications\footnote{In order to estimate the number of interlopers and the fraction of true lines lost in the forest, we have created a large set of artificial spectra. For details see next section.}, and $\Delta z_{tot}${} the total pathlength covered by all spectra used for assembling the sample. 
The error on this redshift density is
\begin{equation}
\sigma _{dn/dz} = \frac{\sqrt{n_{tot} + n_{rand}}}{\Delta z _{tot}} 
\end{equation} 
The total pathlength for a sample can be found by summing up all sightlines that cover various intervals of redshift. In our case, all AGN spectra that were searched for \ion{O}{6}{} absorbers had the same low redshift boundary, defined by the instrumental setup of the SDSS spectroscopy \citep[]{york2000}. In order to avoid including intrinsic and associated absorbers into our sample, we have chosen to define the maximum redshift such that the velocity difference between the AGN's emission and the absorber is 5000 km/s (where we define blue-shifted absorbers to have positive velocity values). This 'arbitrary' velocity cut-off proves to  select against most of the galaxies clustered around the QSO (and ionized by the QSO); all the gas in the $>$ 1 kpc region of the QSO host, and most of the ``intrinsic'' absorbers \citep[]{wild2008}. But note that it has become clear that a small fraction of the ``intrinsic'' absorbers can have $\beta${} up to 12,000 km/s (\citet[]{vandenberk2008} and \citet[]{nestor2008}. Hence, limiting ourselves to the lower cut-off of 5000 km/s will, in the strict sense, provide only upper limits for the number of 'intervening' absorbers above this cut-off\footnote{Note that the opposite case, intervening absorbers at $\beta <$5000 km/s is also possible. Hence, the true total number density of intervening absorbers cannot be assessed here.}, but we estimate that the errors introduced by the unknown fraction of 'intrinsic' absorbers with $\beta \geq${} 5000 km/s are negligible compared to the other systematic and statistical errors of our algorithm.\\
Fig. 1 shows $g(z)$, the number of lines-of-sight within $z${} and $z+dz$, from which the redshift path $\Delta z${} can be easily obtained :
\begin{equation}
\Delta z = \Sigma g(z)\delta z
\end{equation}  
where $\delta z$ is the resolution of the grid. Note that the overwhelming majority of sight-lines in the full sample is comprised of spectra with a signal-to-noise ratio (SNR) within the Lyman forest below the nominal value of 4.0, that an AGN needs to obtain {\em overall}{} to enter the SDSS QSO database. In order to ensure comparability with other studies, and to allow for the unambiguous detection of \ion{O}{6}{} absorption at a reasonable significance level, we have chosen to use a subsample of the full dataset which includes only sources for which SNR$_{forest} > 5.0$. In the following, this is going to be the sample from which we draw inferences unless otherwise stated. It consists of 387 AGN, and allows for the detection of \ion{O}{6}{} absorbers above a rest-frame equivalent width limit of the 1032 \AA{} component of $W_{r} > 0.19 \AA$.   

\begin{figure}
\includegraphics[angle=270,width=\columnwidth]{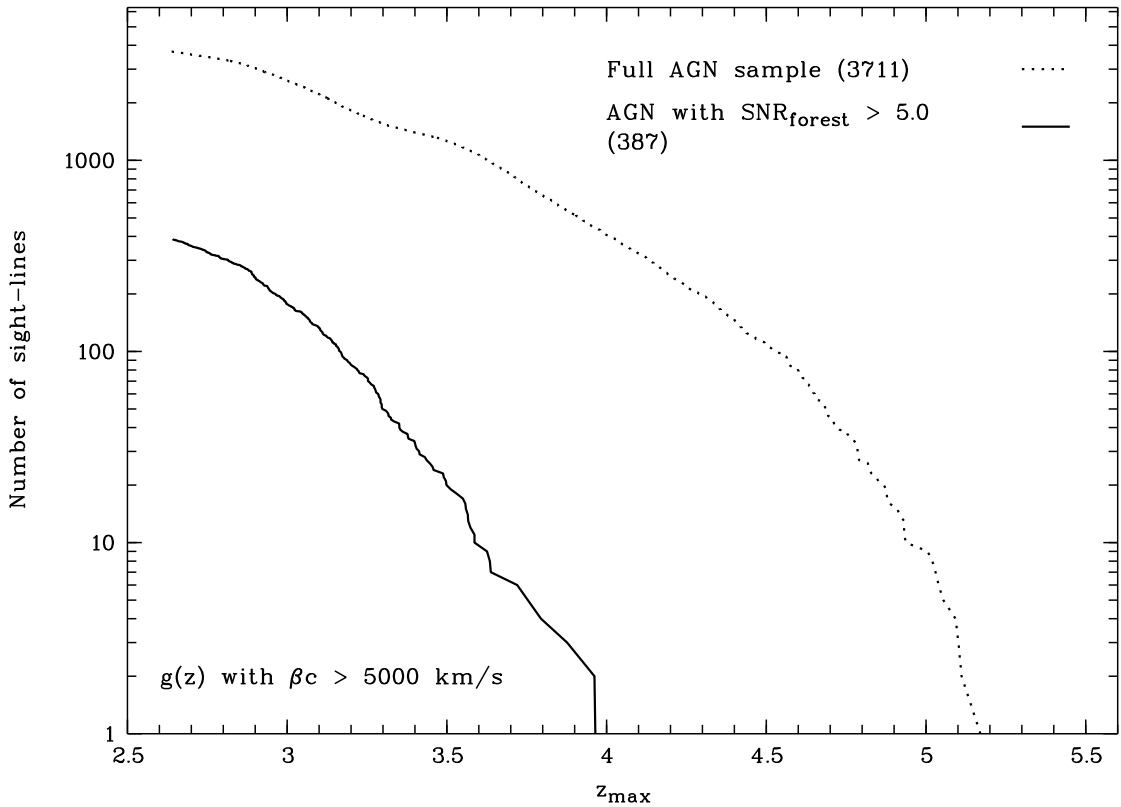}\label{gz} 	
\caption{The number of sightlines towards AGN in the full SDSS sample, and the subset of sources where the SNR$_{forest} >$ 5.0, versus the maximum redshift allowed in the spectrum for detecting intervening absorbers. We have chosen a velocity difference between the absorber's redshift and the quasar emission of $\beta c >$ 5000 km/s.}
\end{figure}

Table 1 lists the number of AGN, the total pathlength, the total number of absorbers found, as well as the estimated number of random absorbers per redshift interval, and the line density derived from these quantities. For comparison it also lists the findings of various other studies for \ion{O}{6}{} and other absorbers. Note that these values represent the {\em observed}{} line densities based upon actual detection of lines, no correction for the fraction assumed to have gone undetected has been applied. Thus, in a strict sense the measurements only provide a secure lower limit.   For the full sample over the complete redshift range $2.8 \leq z_{abs} \leq 5.1$, i.e. including all sources regardless of S/N within the forest, a density $\Delta n/\Delta z = 0.21 \pm 0.01${} is obtained. This is much lower than the value derived by \citet[]{burles1996} for their sample at an average redshift of $z_{abs} = 0.9$, but this can be explained both by the fact that our identification efficiency degrades rapidly towards higher redshifts (as detailed in \citet[]{frank2007}, and that, as noted above, there does not exist a clearly quantifiable equivalent width limit $W_{r,limit}${} for the majority of the sources where SNR$_{forest} \leq 4.0$. When focusing on possible absorbers in the redshift interval $2.8 \leq z_{abs} \leq 3.2${} within the sample of high SNR$_{forest}$, where we expect the detection efficiency to be best, our line density of $\Delta N/\Delta z (2.8 \leq z_{abs} 3.2) = 1.31 \pm 0.12${} is in good agreement with the lower redshift value of \citet[]{burles1996}, indicating that there is no evolution of the rate of incidence of strong \ion{O}{6}{} absorbers over this wide range of redshifts. We stress, however, that although the \citet[]{burles1996}{} analysis was performed with similar spectral resolution and SNR, the fact that the Lyman $\alpha${} forest density is much higher at the redshifts probed in our work, probably leads to a much more severe underestimate of the line density in our case due to the higher fraction of lines lost in the forest by blending with strong \ion{H}{1}{} absorbers, as we will see in the next section.\\
The comparison to searches for \ion{O}{6}{} absorbers at only slightly lower redshifts than probed by our study is difficult, since most of these were performed using high SNR and much higher spectral resolution than the SDSS spectra allowed. Thus, it is not surprising that studies like \citet[]{simcoe2004, bergeron2005, carswell2002}{} find a substantially higher $\Delta n/\Delta z${} at z$_{abs} \sim 2.0 - 2.5$: the limiting equivalent width in their cases drops to $W_{r, limit} \sim 0.05 \AA$, allowing to probe the more abundant weaker systems. Simply dropping their weak absorbers from their samples to derive the frequency of strong \ion{O}{6}{} absorbers does not provide a reliable estimate, because multiple weak and for SDSS unresolvable components of a complex absorbing system can form a feature detectable with our search algorithm, yet the number of lines identified in such cases in the high resoultion studies strongly depends on the often detailed velocity and column density structures of the components. Hence, in a strict sense, we need to compare the number of 'systems', and not the individual components of these studies with our 'candidates' that all represent strong features. Of more importance, however, is the fact that \citet[]{bergeron2005}{} employ a different method in estimating $\Delta n/\Delta z$, they infer the line density by integrating over a fit to the column density distribution f(N): $\Delta n/\Delta z = \Delta X/\Delta z \times \int f(N)dN$, where $\Delta X${} is the redshift path density. This procedure requires detailed knowledge of the column density sampled by each absorber, and can usually only be obtained with high resolution and high S/N, as we will detail in the next section.  

\begin{deluxetable}{ccccccccc}\label{line_density}
\tablecaption{Observed Line Density Of Various Absorption Line Studies}
%\tablehead{
%\colhead{Ion} &
%\colhead{$z_{abs}$} &
%\colhead{Sample size} &
%\colhead{$\Delta z_{tot}$} &
%\colhead{N$_{tot}$} &
%\colhead{N$_{rand}$} &
%\colhead{$\Delta N/\delta z$} & 
%\colhead{Reference}

\startdata
Ion & $z_{abs}$ & Sight-lines & W$_{r, limit}$ & $\Delta z_{tot}$ & n$_{tot}$ &  n$_{rand}$ & $\Delta n/\Delta z$ & Reference \\
\hline  
\ion{O}{6}	&  2.8 - 3.2 & 305 & 0.19 \AA & 98.9 & 139 & 9.8 & 1.31 $\pm$ 0.12 & 1 \\
\ion{O}{6}	&  3.2 - 3.6 & 84  &  0.19 \AA & 53.2 & 41  & 3.5 & 0.71 $\pm$ 0.14 & 1 \\
\ion{O}{6}	&  3.6 - 4.0 & 9  &  0.19 \AA & 9.74 & 5  & 0.63 & 0.45 $\pm$ 0.24 & 1 \\
\hline
\ion{O}{6}	&  2.8 - 5.1 & 3366  &  -- & 3768.4 & 1290  & 508.9 & 0.21 $\pm$ 0.01 & 2 \\
\hline 
\ion{O}{6}	&  0.9 & 11  &  0.21 \AA & 4.73 & 6  & 1.3 & 1.0 $\pm$ 0.6 & 3 \\
%\ion{Mg}{2}	&  0.9 & ?  &  0.30 \AA & ? & ?  & ? & 1.0 $\pm$ 0.25 & 4 \\
%\hline
\ion{O}{6}	&  2.2 - 2.5 & 7  &  0.04 \AA & 2.016 & 230 *  & n.a. & 114  & 4 \\
\ion{O}{6}	&  2.2 - 2.5 & 7  &  0.04 \AA & 2.016 & $\sim$100 **  & n.a. & 50 $\pm$ 5  & 4 \\
\hline
\ion{O}{6}	&  1.99 - 2.57 & 10  &  ? \AA &  & 136  & n.a. & 74 (66..106) &  5 \\
\ion{O}{6}	&  1.99 - 2.57 & 10  &  0.05 \AA & ? & ?  & n.a. & 26 &  5 \\
\hline
%\ion{O}{6}	&  0.1 (range?) & ?  &  0.05 \AA & ? & ?  & ? & 13 &  7 \\
%\hline
%\ion{O}{6}	&  0.17 - 0.27 & ?  &  0.03 \AA & ? & ?  & ? & 48 &  8 \\
%\hline
\ion{O}{6}      &  $<0.5$  & 16 &  0.03 \AA & 2.623 & 55 & n.a. & 21.0$\pm$3.0 & 6 \\
\ion{O}{6}      &  $<0.5$  & 16 &  0.30 \AA & 3.179 & 1  & n.a. & 0.3 +0.7/-0.3 & 6 \\
%\ion{O}{6}	&  2.1-2.3 & 2  &  0.03 \AA & ? & ?  & ? & 5.7 $\pm$ 2.9  &  7 \\
\ion{O}{6}	&  2.1-2.3 & 2  &  0.03 \AA & 0.70  & 8 & n.a. & 11.4 $\pm$ 4.0 &  7 \\
\enddata
%\tablenotetext{a}{to be added}
\tablerefs{Note that lines density, as used by the authors, refers more strictly to components of systems, which are all counted as separate entities.\\
           1. This work, SNR$>$5 sample,\\
           2. This work, full sample, W$_{r, limit}$ not constant.,\\
           3. \citet[]{burles1996},\\
           %4. \citet[]{fan1995},\\
           4. \citet[]{simcoe2004}{} - high SNR + Resolution, slightly diffferent method (survival analysis), * including also upper limits, ** including detected lines only\\
           5. \citet[]{bergeron2005}{} - high SNR + Resolution, from fit to CDDF,\\
           no EW limit given for first value, in second row no values given for number of lines and pathlength\\
           %7. \citet[]{sembach2004}{} - FUSE/HST \\
           %8. \citet[]{tripp2000}\\
           6. \citet[]{tripp2008}, no correction for false identifications
           7. \citet[]{carswell2002}. }
\end{deluxetable}

%%%%%%%%%%%%%%%%%%%%%%%%%%%%%%%%%%%%%%%%%%%%%%%%%

\subsection{The incompleteness and S/N corrected Redshift Number Density}
As indicated in the preceding section, the 'raw' number density obtained by simply counting the \ion{O}{6}{} candidates found with our search algorithm has to be modified in order to find the 'true' number of such absorbers. As a first step, we have already incorporated an estimate of the number of spurious \ion{H}{1}{} interlopers that mimick \ion{O}{6}{} doublets. This estimate stems from running our search algorithm on all spectra with a wavelength separation between the two putative absorption features that has a random offset from the nominal rest-frame $\Delta \lambda =  1037.613 \AA - 1031.912 \AA = 5.701 \AA$, as detailed in \citet[]{frank2007}. Far more important, however, is the estimate of fraction of candidates lost through blends with ubiquitious Lyman forest lines. For each candidate we have calculated this fraction by creating a set of mock forest spectra, as we describe below.\\
We take the absorber redshift, the equivalent width estimate, the AGN emission redshift and the S/N measurement inside the relevant part of the Lyman forest from our catalogue list of candidates \citep[]{frank2007}. Then we create 200 different line lists, randomly populated with \ion{H}{1}{} absorbers according to the line density estimate of \citet[]{kim2001}{} with a column density distribution following \citet[]{hu1995}, for the appropriate emission redshift z$_{em}${} covering the complete wavelength range from the SDSS low-cutoff ($\sim$ 3800 \AA) up to the Lyman $\alpha${} emission of the quasar. From this list we create a mock high-resolution spectrum by modelling the Lyman absorbers with Voigt profiles, where the Doppler parameter $b${} is drawn from a Gaussian distribution with a mean of $b_{c} = 25.0${} km/s and width of $\Delta$ = 8 km/s. We have modelled all \ion{H}{1}{} transitions down to Ly $\epsilon$. We then degrade the spectrum to the SDSS resolution, add the appropriate level of noise to each pixel, and place the \ion{O}{6}{} candidate into this model of the Lyman forest. Running the same detection algorithm on these spectra as used in \citet[]{frank2007}, we can thus estimate how often the Lyman forest destroys the signal. The fraction f$_{i}${} of absorbers retrievable is therefore a function of the absorber redshift, the emission redshift of the quasar, the average S/N within the forest, and the strength of the absorber as measured by its equivalent width :
\begin{equation}
f_{i} = f_{i} (z_{abs},z_{em}, S/N, W_{r} (1032 \AA), W_{r} (1038 \AA))
\end{equation} 
We can then proceed to estimate the 'true' number $n_{tot, c}${} of absorbers that we should have found by summing over all candidates $i$ :
\begin{equation}
n_{tot, c} (W_{r}>W_{r, limit}) = \sum _{i} \frac{1}{f _{i}}
\end{equation}
Exactly the same procedure can be performed for the estimate of the number of random interlopers :
\begin{equation}
n_{rand, c} (W_{r}>W_{r, limit}) = \sum _{j} \frac{1}{f_{j}}
\end{equation}
where the summation is over all $j${} candidates found via the 'random' search.\\
Additionally, we have taken into account that there are individual pixels in the spectra that will fall below the S/N limit we have imposed in the first step for the {\it average}{} S/N inside the Lyman forest. We thus have to reduce each individual redshift path $\Delta z _{i}${} towards the $i$th quasar by removing such 'bad' pixels. This results in a lower overall $\Delta z _{tot,c}${} for each redshift interval of interest. This effect is much more pronounced towards higher redshifts, as the rapidly increasing density of the Lyman forest introduces more and more regions with substantial flux decrements, leading to lower individual S/N at the affected wavelengths.\\
It is interesting to see that these corrections reverse the trend within the redshift bins available to us of a decreasing number density with increasing redshift : while we now arrive at $\Delta n_{c}/\Delta z _{c} (2.8 \leq z_{abs} \leq 3.2) = 4.6 \pm 0.3$, the values for $3.2 \leq z_{abs} \leq 3.6$, and $3.6 \leq z_{abs} \leq 4.0${} increase to $\Delta n_{c}/\Delta z _{c} = 6.7 \pm 0.8$, and $8.4\pm2.9$, respectively. As Table 2 shows this is mainly due to the vastly decreased corrected redshift path length. Due to the large errors, and the risk of being susceptible to cosmic variance with such short pathlengths, we cannot assess the reality of this trend. If, however, this trend was real, then it would imply an extremely strong evolution of the redshift path  density : $\Delta n_{c}/\Delta z _{c} \sim (1+z)^{\gamma}$, where $\gamma \sim 2.5-3.0$, comparable or even stronger than the evolution of e.g. strong \ion{Mg}{2}{} absorbers at $z_{abs} \leq 2.3$ seen by \citet[]{steidel1992} or \citet[]{prochter2006}. A comparison to the studies mentioned in the preceding section is not possible, because none of these has incorporated all the corrections we have applied here, and also probably sample different absorbers, as {\bf we indicate in the next section}.  
\begin{deluxetable}{cccccccc}\label{line_density_corrected}
\tablecaption{Incompleteness and S/N corrected Line Density Estimates}

\startdata
Ion & $z_{abs}$ & Sight-lines & W$_{r, limit}$ & $\Delta z_{tot,c}$ & n$_{tot,c}$ &  n$_{rand,c}$ & $\Delta n_{c}/\Delta z _{c}$ \\
\hline  
\ion{O}{6}	&  2.8 - 3.2 & 305 & 0.19 \AA & 50.5 & 251.7 & 19.4 & 4.6 $\pm$ 0.3 \\
\ion{O}{6}	&  3.2 - 3.6 & 84  &  0.19 \AA & 12.3 & 92.7  & 10.3 & 6.7 $\pm$ 0.8 \\
\ion{O}{6}	&  3.6 - 4.0 & 9  &  0.19 \AA & 1.37 & 13.75  & 2.2 & 8.4 $\pm$ 2.9 \\
\hline
\enddata

\end{deluxetable}

\begin{figure}
\includegraphics[angle=270,width=\columnwidth]{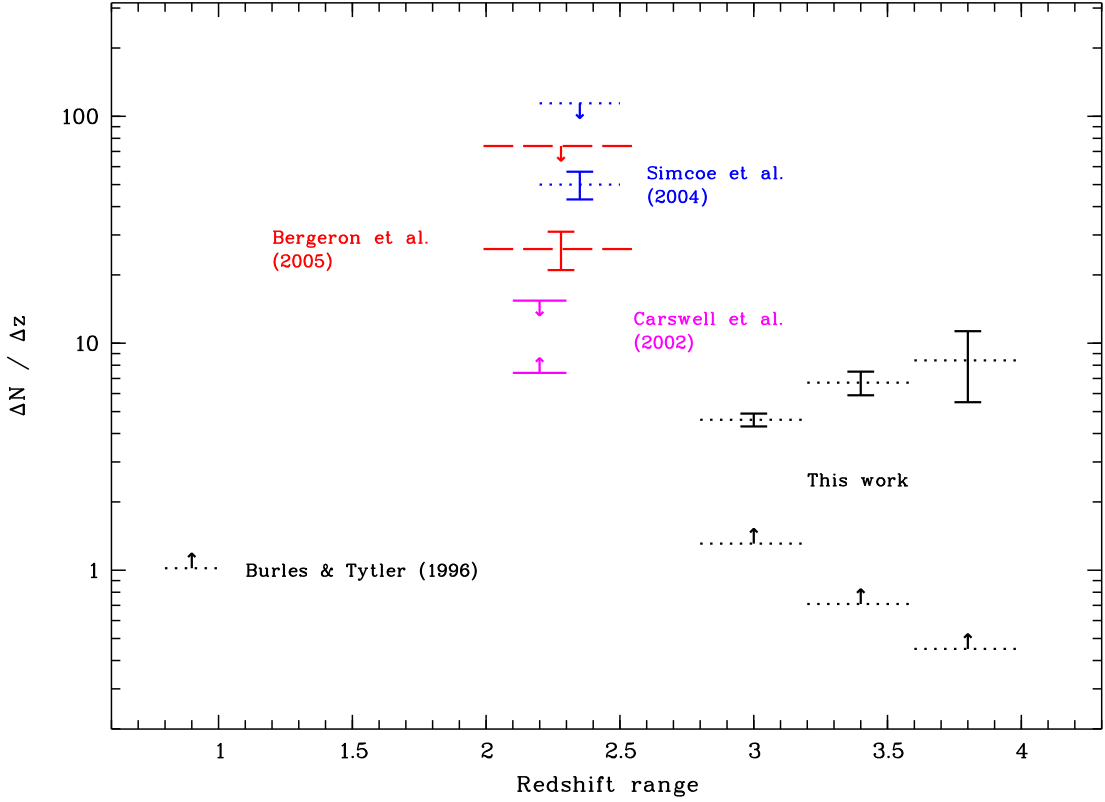}\label{dn_dz} 	
\caption{The {\bf OVI}{} line density estimates for our and other studies. Our data are represented by the three redshifts bins above $z_{abs} \geq 2.8$. Indicated are the secure lower limits on $\Delta n / \Delta z${} that we can place simply by directly counting our candidates and summing up all redshift paths (arrows). By accounting for the losses of absorbers in the dense forest via analysing a large set of mock spectra, and the reduced pathlength induced by the noise in certain pixels being higher than allowed for our detection limit of $W_{r} \geq 0.19 \AA$, we are able to derive the {\it corrected} values of $\Delta n / \Delta z${} shown for the same bins (points with error bars). The measurements and/or upper and lower limits for other studies are labeled, and their redshift range is indicated by the length of the bins. Note that the three studies at intermediate redshifts ($2.0 \leq z \leq 2.7$) are performed with high-resolution spectroscopy (HIRES or UVES), while the \citet[]{burles1996} is done with similar resolution as ours (FOS/HST).  {\bf The upper limit of dn/dz = 114 indicated for the Simcoe sample (blue dotted lines) includes all of their upper limits/non-detections. The data point for Bergeron (red dashed) is their conservative estimate when applying a rest-frame equivalent width limit of 50m\AA, dropping this limit they obtain dn/dz=74(+32-6), as indicated by the dashed upper limit.} For details see text. }
\end{figure}

%%%%%%%%%%%%%%%%%%%%%%%%%%%%%%%%%%%%%%%%%%%%%%%%%
%%%%%%%%%%%%%%%%%%%%%%%%%%%%%%%%%%%%%%%%%%%%%%%%%

\section{The Cosmological Mass Density}

%%%%%%%%%%%%%%%%%%%%%%%%%%%%%%%%%%%%%%%%%%%%%%%%%
%%%%%%%%%%%%%%%%%%%%%%%%%%%%%%%%%%%%%%%%%%%%%%%%%
 
The contribution of a certain trace element in a particular ionisation state to the current closure density of the universe can be estimated from the measurement of the column densities (Lanzetta et al. 1991, Songaila et al. 2001). In our case :
\begin{equation}
\Omega _{OVI} = \frac{1}{\rho _{c}} m_{OVI} \frac{\sum_{i} N_{OVI, i}}{c/H_{0} \sum_{i} \Delta X_{i}}
\end{equation}
where $\rho_{c} = 1.89 \times 10^{-29}$ g cm$^{-3}${} is the current critical density, $m_{OVI}$ the mass of the \ion{O}{6} ion, $H_{0}$ the current Hubble parameter, and $N_{OVI,i}${} the total \ion{O}{6}{} column density toward the $i$th QSO. The absorption path distance $\Delta$X between the maximum and minimum redshifts of a sight-line towards a QSO, $[z_{min},z_{max}]$, for a flat universe is defined as
\begin{equation}
\Delta X = {[\Omega _{m}(1+z_{max})^{3} + \Omega_{\lambda}]^{1/2} - [\Omega _{m}(1+z_{min})^{3} + \Omega_{\lambda}]^{1/2}}
\end{equation}
The standard error of this mass density can be estimated to :
\begin{equation}
\sigma _{\Omega} = \frac{1}{[1-(1/n)]^{0.5}}\frac{\sqrt{\sum_{i} [N_{OVI, i} - <N_{OVI}>]^{2}}}{\Delta X_{tot}}
\end{equation}
where $n$ is the total number of absorption systems in the sample, and $<N_{OVI}>${} is the mean column density of the sample (Burles 1996).

\begin{figure}
\includegraphics[angle=270,width=\columnwidth]{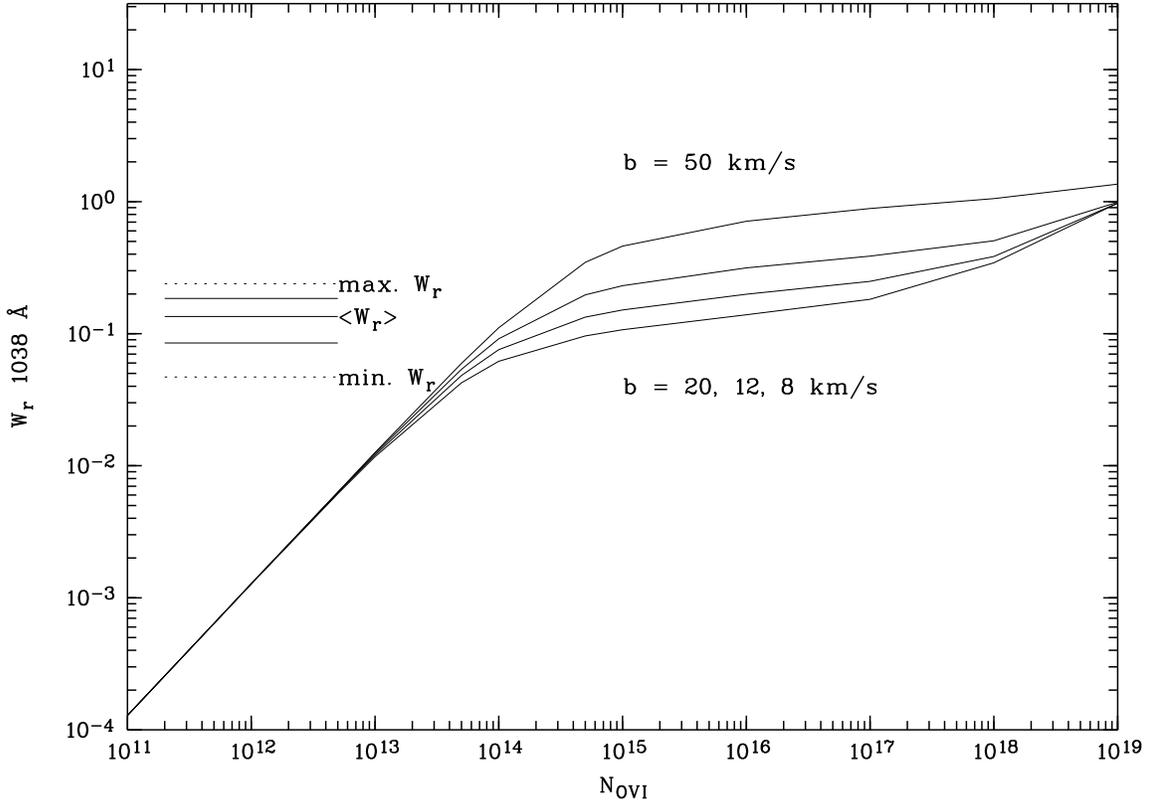}\label{curve_of_growth} 	
\caption{The curve of growth for \ion{OVI} 1038 \AA. The relation between the rest-frame equivalent width W$_r${} and the column density N$_{OVI}${} is obtained by modelling the 1038 \AA{} transition with a Voigt profile with a given Doppler parameter $b$. Note that the linear part the curve of growth does not depend on the choice of $b$, whereas the flat part retains that model dependency. {\bf The solid lines on the left side show the 25th percentile, average, and 75th percentile, while the dotted lines indicate the  extreme values of the equivalent width distribution of the \ion{O}{6}{} lines in the SNR$>$5 sample.} }
\end{figure}

A lower limit on the column density $N_{OVI}${} in each absorbing system can be obtained by measuring the rest-frame equivalent width of the weaker 1038 \AA{} transition $W_{r}$(1038 \AA). Fig. \ref{curve_of_growth}{} shows the curve-of-growth for this ion under various model assumptions. If the absorption features are unsaturated, the column density and rest-frame equivalent width are simply proportional to each other. For saturated lines, however, the relation retains a model dependency : larger broadening parameters $b${} lead to larger $W_{r}$. In a first step, we have extracted $W_{r}$(1038 \AA) from the calalogue of all lines identified by our pixel-by-pixel search algorithm (Frank et al. (2010) in the high SNR sample of spectra (SNR$_{forest} >$5.0, 387 spectra, 237 \ion{O}{6}{} identifications). In order to decide whether two absorption features at the correct wavelength separation can qualify as \ion{O}{6}{} 1032/1038 \AA{} doublet, the automatic search program 'measures' the equivalent width of an absorption feature simply by multiplying the flux decrement in one pixel with the pixel wavelength separation. This gives, as detailed in Frank et al. (2007), a lower limit to the equivalent width as an absorption feature might span more than one pixel. In Fig. \ref{curve_of_growth}{} the location of the average $W_{r}$(1038 \AA) and the extreme values found in our sample are denoted by the solid and dotted lines. It is obvious that even these lower limits for the equivalent widths begin to populate the regime where the reliable estimate of the column density from $W_{r}${} alone is not possible. We can, however, draw sound inferences on lower limits : Fig. \ref{curve_of_growth}{} shows that even for unreasonably high $b${} paramaters (note that \citet{simcoe2002}{} obtain a median $b = 16${} km/s) very few of the absorbers can reside below $N_{OVI} < 10^{14}${} cm$^{-2}$, and the average for the sample is well above that limit. Thus, we can safely assume that $\sum_{i} N_{OVI, i} \geq i \times 10^{14}${} cm$^{-2}$.

\begin{deluxetable}{cccccc}\label{mass density}
\tablecaption{Cosmological Mass Density of Various Absorption Line Studies}
%\tablehead{
%\colhead{Ion} &
%\colhead{$z_{abs}$} &
%\colhead{Sample size} &
%\colhead{$\Delta z_{tot}$} &
%\colhead{N$_{tot}$} &
%\colhead{N$_{rand}$} &
%\colhead{$\Delta N/\delta z$} & 
%\colhead{Reference}

\startdata
Ion & $z_{abs}$ & Sight-lines & $\Delta X_{tot}$ & $\Omega \times h^{-1} \times 10^{-8}$ & Reference \\
\hline  
\ion{O}{6}	&  2.8 - 3.2 & 305 & 113.9 & $\geq 1.9$ & 1 \\
\ion{O}{6}	&  3.2 - 3.6 & 84 & 24.1 & $\geq 2.6 $  & 1 \\
\ion{O}{6}	&  3.6 - 4.0 & 9 & 2.6 & $\geq 2.9 $  & 1 \\
\hline
\ion{O}{6}	&  0.9 & 11 & 6.6 & $\geq 14.0 \pm$ 4.0 (q$_{0}$ = 0.5) & 2 \\
\ion{O}{6}	&  0.9 & 11 & 6.6 & $\geq 9.8 \pm$ 3.0 (q$_{0}$ = 0.0) & 2 \\
\hline
\ion{O}{6}	&  2.5  & 7 & ? &  7.89  & 3 \\
\hline
\ion{C}{4}      & 2.5   & 7 & 3.27 & 1.78  & 3 \\ 
\hline
\ion{O}{6}	&  2.3 & 10 & ? &  15.1  & 4 \\
\ion{O}{6}	&  2.3 & 10 & ? &  35(26..67)  & 4 (fit) \\
\enddata                                                                                                                                                                                                     

%\tablenotetext{a}{to be added}
\tablerefs{1. This work, SNR$>$5 sample,
           2. Burles \& Tytler 1996 - note different cosmology !,
           3. Simcoe 2004, using Survival method.
           4. Bergeron \& Herbert-Fort (2005), using a fit to the column density distribution, and two different sample criteria}
\end{deluxetable}

\begin{figure}
\includegraphics[angle=270,width=\columnwidth]{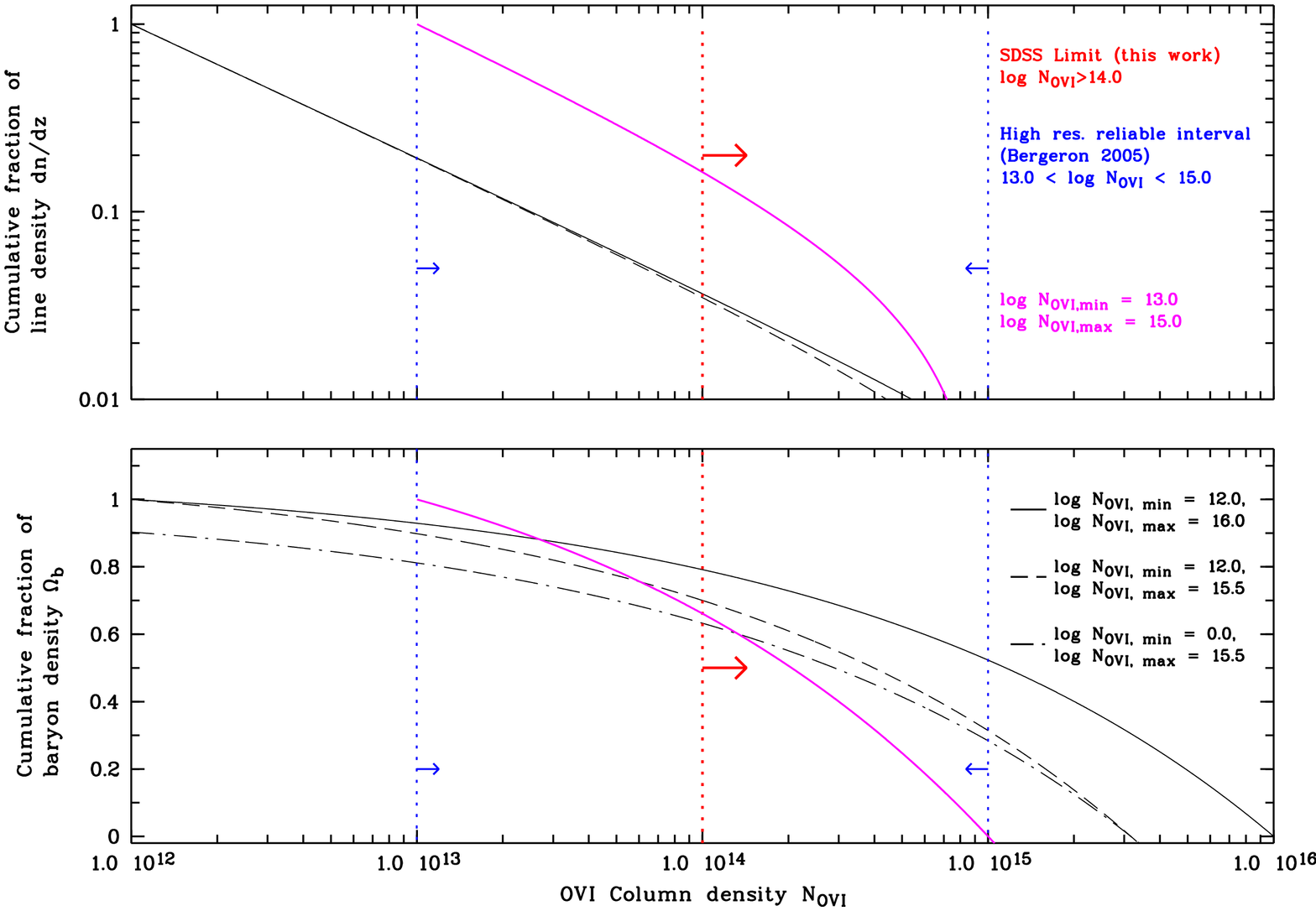}\label{cumul_dist} 	
\caption{The cumulative fractions of the line density dn/dz (upper panel) and the contribution to the baryon density $\Omega _{b}${} of \ion{O}{6}{} absorbers in the IGM, assuming a power-law distribution for the column density f(N) $\sim N^{-\alpha}$(\citet{bergeron2005}, $\alpha = 1.71$. The red and blue dotted lines indicate the limits of our and the \citet[]{bergeron2005}{} analyses, respectively. Note that while we do not retrieve a substantial fraction of all lines, we nevertheless sample $\geq 65\%${} of the mass in such absorbers. For details see text.}
\end{figure}

Table \ref{mass density}{} shows the results of these calculations. For the redshift bin with the highest confidence and the best success rate of the automatic search algorithm, we obtain a lower limit for the mass density of $\Omega _{OVI} (2.8 < z_{abs} < 3.2) \geq 1.9 \times 10^{-8} h^{-1}$. 
How does this value compare to measurements at lower redshifts ? Burles \&{} Tytler (1996), analysing FOS spectra with similar resolution, arrive at an \ion{O}{6}{} density limit at an average redshift of z$_{ave} = 0.9$ of $\Omega _{OVI} \geq 9.8 \pm 3.0 \times 10^{-8} h^{-1}$ (but note that they use a different cosmology). The inspection of two QSO sightlines with high resolution UVES spectra by Carswell, Schaye and Kim (2002) yields $\Omega _{OVI} = 6 \times 10^{-8} h^{-1}$ at $z \sim 2.2$, and the analysis of Simcoe et al. (2004) for a sample of $z_{abs} \sim 2.5$ absorbers gives $\Omega _{OVI} = 7.89 \times 10^{-8} \times h^{-1}$. {\bf Note that \citet[]{bergeron2005} give a much higher mass density over the range 2.0 $\leq z_{abs} \leq$ 2.6, when accounting for incompleteness via integrating over the column density distribution function : $\Omega _{OVI} = 24.5 \times 10^{-8} \times h^{-1}$. Hence, there is a wide range of values found by different authors using different methods at slightly lower redshifts than probed by us.} Apart from the Burles \&{} Tytler (1996) study, these values are derived from analyses of high-resolution and high-S/N data. And yet our estimate based upon SDSS spectra already allows us to pose a lower limit at a higher redshift that lies only a factor of 3-5 below {\bf the mass density esimates of Carswell, Schaye and Kim (2002), and the one of  Simcoe et al. (2004). It is interesting to note that the incompleteness correction terms introduced for our line density estimate are of that order. We remain, however, a factor of 13 lower than \citet[]{bergeron2005}, and hence whether there is a trend suggesting for the \ion{O}{6}{} mass density to decrease when moving from $z_{abs} \sim 2.5${} to $z_{abs} \sim 3.2$, is thus not completely clear. This also means that the more abundant, yet weaker \ion{O}{6}{} absorbers probed by the Simcoe et al. (2004) study cannot contribute much to the \ion{O}{6}{} mass density}.\footnote{Note that these authors also specifically make an incompleteness correction by integrating over a column density distribution function and a metallicity function, that was obtained by a survival analysis taking into account non-detections.}\\
Indeed, a closer examination of the cumulative fractions of the line density and the cumulative contribution to the baryon density of the absorbers found in our SDSS search reveals that we sample already a high fraction of the mass, despite missing a large fraction of the lines as we cannot detect weak absorbers. Figure \ref{cumul_dist}{} shows the fraction of \ion{O}{6}{} absorbers detected as a function of the lower limit to the column density sampled (upper panel). This cumulative fraction, and the contribution to the baryon density $\Omega _{b}${} of these absorbers (lower panel), are derived assuming a power-law dependence for the column density distribution function \citep[]{bergeron2005} : $f_{N_{OVI}} \sim N_{OVI} ^{- \alpha}$. The line density per redshift interval, dn/dz, is then obtained by simply integrating the column density function :
\begin{equation}
dn/dz (N_{low} \leq N_{OVI} \leq N_{up}) =  \int \limits_{N_{low}}^{N_{up}} f(N)dN \sim N ^{-\alpha+1}
\end{equation} 
where the upper boundary for the \ion{O}{6} column density, $N_{up}$, for the absorbers can be estimated from theoretical models for the IGM or assumed from published lists, and the lower boundary, $N_{low}$, is in principle a function of the detectability limit, varying with instrumental set-up and the redshift intervals probed. Here we have chosen to adopt the slope $\alpha = 1.71${}, found by \citet[]{bergeron2005} for redshifts $z_{abs} \sim 2.5$, assuming that there is no substantial redshift evolution. For the upper limit to the column density, we apply two different values. We have checked the published lists of \ion{O}{6}{} absorbers, and found that there are only very few examples of lines as strong as log $N_{OVI} \geq 15.5$. Even for damped-Lyman $\alpha${} (DLA) absorbers with their high column density for neutral hydrogen, the \ion{O}{6}{} absorbers associated with them tend to fall below this limit \citet{fox2006}. Hence, we limit ourselves in the first case to column densities below $N_{OVI} = 10^{15.5}$. A second, sensible limit can be derived by integrating the \citet[]{bergeron2005}{} column density distribution, and calculating the number of strong absorbers we expect to be able to retrieve within our given pathlength. Beyond column densities of log $N_{OVI} \geq 16.0$, this number drops below three absorbers in the redshift range $2.8 \leq z_{abs} \leq 3.2$, and below one at higher redshift. We have thus chosen log $N_{OVI} = 16.0$ as our limiting column density for the second case. We stress that these upper limits probably reflect a physical boundary : when the gas density increases beyond a peak value for each species, the recombination drives the ionisation equilibrium fraction for the \ion{O}{6} ionic species to lower values than can be compensated for by the increasing gas density, and hence the \ion{O}{6}{} density will also drop (in other words : recombination is dependent on $n_{e}^{2}$, whereas ionisation only depends linearly on $n_{e}$).\footnote{Note that this is not true for the case of collisional ionisation.} There is no physical limit for the lower boundary, but lines too weak simply become undetectable. A column density of $log N_{OVI} = 12.0${} produces an absorption feature that is right at the detection limit of typical high-resolution and high S/N exposures with 8m-class telescopes and instruments like UVES or HIRES. Hence, we have set this value as our lower boundary $N_{low}$. Note that the lower boundary is much more important for the fraction of lines that can be discovered, than for the cumulative mass fraction sampled, as the power-law index $\alpha = 1.71${} is close to 2.\\ 
The baryonic mass traced by all detected lines within the interval given by the two boundaries discussed above, is proportional to the intergral
\begin{equation}
\Omega _{b} (N_{low} \leq N_{OVI} \leq N_{up}) \sim \int \limits_{N_{low}}^{N_{up}} Nf(N)dN \sim N ^{-\alpha+2}
\end{equation}
Figure \ref{cumul_dist}{} shows in the upper panel the cumulative fraction of the line density as a function of the limiting column density in a survey. The red dotted line and arrow indicate how low our SDSS based sample reaches, and the interval indicated by the blue dotted lines and the arrows delineates the boundaries chosen by \citep[]{bergeron2005}{} to maximise completeness and minimise the effects of cosmic variance in their sample of UVES spectra. Note that our long effective pathlength allows us in principle to probe out to higher column density systems without being afflicted by the rarity of such absorbers causing problems with cosmic variance. Clearly, we cannot expect to sample more than 5\%{} of the lines that can be retrieved with typical high-resolution and high-S/N observations.  As the lower panel shows, however, we are picking up the bulk of the mass density of \ion{O}{6}{} absorbers with our approach : depending on the exact upper limit to the column density of such absorbers, we are sensitive to  at least 65\%{} of the absorbers by mass, and assuming the same lower cut-off as for the number density (log N$_{OVI} \geq 12.0$), we {\bf are in principle able to}{} sample up to 82\%{} of the mass, {\bf not accounting for the inevitable incompleteness of the survey towards the lower end of the column density threshold.}\footnote{While the fraction of lines that can be retrieved is a strong function of the integration limits for the column density, the fraction of the mass sampled depends mostly on the high-cutoff value. But even restricting ourselves to the regime between $13.0 \leq log N_{OVI} \leq 15.0${} means that we are probing a high fraction of the mass, as the magenta curves in this figure demonstrate.}        

%%%%%%%%%%%%%%%%%%%%%%%%%%%%%%%%%%%%%%%%%
%%%%%%%%%%%%%%%%%%%%%%%%%%%%%%%%%%%%%%%%%

\section{A lower limit for the Cosmic Metallicity}
From the cosmological mass density at redshift $z$, it is a straightforward process to estimate the mean metallicity of the gas due to OVI. Following \citet[]{burles1996}{}, let $\zeta (z)${} be the ratio of mean metallicity\footnote{Note that this definition takes into account only the gas containing strong OVI systems accompanied by HI. In addition, it implicitly assumes a volume filling factor of one for the gas with such detected systems in SDSS spectra. Given these caveats, it is still a safe lower limit that we derive.}{} at redshift $z${} to the solar value, then
\begin{equation}\label{mean_metallicity}
\zeta (z) = \frac{\mu}{16} (\frac{H}{O})_{Solar} \frac{O}{OVI} \frac{\Omega _{OVI}(z)}{\Omega _{b}}
\end{equation} 
where $\mu${} is the mean molecular weight ($\sim 1.3${} for a 75\%{} hydrogen and 25\%{} helium mix by mass), and $(\frac{H}{O})_{Solar} = 1174.9$ \citep[]{anders1989}. The cosmological baryon mass density as measured by \citep[]{tytler1994} from high-z deuterium is $\Omega _{b} = 0.023 h^{-2}$. With the current data, the ionisation fraction $\frac{O}{OVI}${} cannot be determined, but we are able to place a strict upper limit for the \ion{O}{6}{} fraction of 22\%, both for collisionally excited or photoionised gas \citep[]{aguirre2007}.\\
With the lower limit for the \ion{O}{6}{} mass density from above, we obtain 
\begin{equation}\label{limit_for_metallicity}
\zeta (2.8 \leq z \leq 3.2) \geq 3.6 \times 10^{-4} h
\end{equation}
or expressed in the usual logarithmic units,
\begin{equation}\label{limit_for_metallicity_log}
[O/H] \geq -3.6
\end{equation}
Burles and Tytler (1996 and erratum 2002) derive $\zeta (z_{abs} \sim 0.9) \geq 4.0 \pm 1.2 \times 10^{-4} \times h$ {\bf , but note that these authors operate with an ionization fraction of OVI/O = 1. Adjusting to our our correction factor, their result ($\zeta (z_{abs} \sim 0.9) \geq 18.2 \pm 5.5 \times 10^{-4} \times h$) is a factor of 5 higher than our (not very constraining) lower limit.}. Carswell, Schaye and Kim (2002) obtain  $\zeta (z_{abs} \sim 2.2) \sim 16 \times 10^{-4} \times h$, whereas Simcoe et al. (2004) propose a median [O/H] = -2.82 (i.e. $\zeta (z \sim 2.5) = 22 \times 10^{-4} \times h$) for the cosmological filaments enriched with oxygen and carbon, but also see examples of systems with very low abundances. They estimate that up to 30\%{} of the Lyman forest lines have [O/H] $\leq -3.5$, i.e. $\zeta \leq 4.5 \times 10^{-4} \times h$. {\bf Using the novel Locally Calibrated Pixel (LCP) search method in $>$2000 SDSS QSOs, \citet[]{pieri2010} find that the results of the search for OVI allows for tightly constraining the {\it mass-weighted} mean oxygen abundance for the IGM ([$<$O/H$>_{MW}$] = -2.33 $\pm$ 0.23 (2$\sigma$), while the {\it volume-weighted} mean abundance is less well constrained : [$<$O/H$>_{VW}$] = -2.74 $\pm$ 0.64 (2$\sigma$). Note, however, that this technique is specifically tailored to the low density regime of the IGM, whereas our approach is complementary in the sense that we are dealing with strong absorbers arising presumably in much denser structures than Pieri et al. (2010) probe.}    
%%%%%%%%%%%%%%%%%%%%%%%%%%%%%%%%%%%%%%%%%
%%%%%%%%%%%%%%%%%%%%%%%%%%%%%%%%%%%%%%%%%

\section{Summary and Conclusions}
We have analysed the \citet[]{frank2007}{} data set of \ion{O}{6}{} absorber candidates found in the spectra of 3800 SDSS quasars, focusing on a subsample of 387 AGN with an average S/N$\geq 5.0${} per pixel inside the Lyman forest region, allowing for the detection of absorbers above a rest-frame equivalent width limit of W$_{r} \geq 0.19 \AA${} for the \ion{O}{6}{} 1032 \AA{} component.\\
Accounting for random interlopers mimicking an \ion{O}{6}{} doublet, we derive for the first time for redshifts $z_{abs} \geq 2.8${} a secure lower limit for the redshift number density $\Delta n / \Delta z$ :\\
$\Delta n _{OVI} / \Delta z (2.8 < z < 3.2) \geq 1.31 $\\
$\Delta n _{OVI} / \Delta z (3.2 < z < 3.6) \geq 0.71 $\\
$\Delta n _{OVI} / \Delta z (3.6 < z < 4.0) \geq 0.45 $\\
In order to quantify the losses of absorbers due to blending with the ubiquitous Ly$\alpha${} forest lines, we have modelled typical Lyman forests following canonical line density and column distribution functions, and estimate the success rate of retrieving each individual candidate as a function of its redshift, the emission redshift of the quasar, the strength of the absorber and the measured S/N for the detection spectrum. Furthermore, we have excluded the parts of the redshift pathlength in each spectrum that fall below the S/N which allows for the detection of lines above the rest-frame W$_{r} = 0.19 \AA$. These correction factors allow us to derive the 'incompleteness and S/N corrected' redshift number densities of \ion{O}{6} absorbers :\\
$\Delta n _{OVI, c} / \Delta z_{c} (2.8 < z < 3.2) = 4.6 \pm 0.3$\\
$\Delta n _{OVI, c} / \Delta z_{c} (3.2 < z < 3.6) = 6.7 \pm 0.8$\\
$\Delta n _{OVI, c} / \Delta z_{c} (3.6 < z < 4.0) = 8.4 \pm 2.9$\\
These values are lower than the ones found in other studies at redshifts $2 \leq z_{abs} \leq 2.5$, owing to the fact that the low SDSS resolution does not allow us to probe systems as weak as in those high-resolution and high S/N studies. Simply comparing these studies with our approach by dropping the weak lines out of the statistics for the high-resolution is not possible : multiple weak and thus for SDSS unresolvable components of a complex absorber can form features that will be detected by our algorithm, yet the number of lines identified in such cases strongly depends on the details of the velocity and column density structures.\\
Transforming the measured equivalent width of the \ion{O}{6}{} candidates into a column density is not possible with the limited information obtainable with the SDSS resolution. We can, however, place a secure lower limit for the detectable \ion{O}{6} column density : log $N_{OVI} \geq 14.0$. With this limit, we can put a constraint on the contribution of \ion{O}{6}{} to the closure mass density at the redshifts probed here:\\
$\Omega _{OVI} (2.8 < z < 3.2) \geq 1.9 \times 10^{-8} h^{-1}$\\
Interestingly, this limit is only a factor of 2-3 below the density estimates of the high-resolution studies mentioned above. Noting that the correction factors due to blending with Ly$\alpha${} and due to the decrease of the available path because of noise are roughly of the same order, this indicates that the weak \ion{O}{6}{} absorbers do not contribute significantly to the mass density. Indeed, we have shown that the strong \ion{O}{6}{} systems that we probe account for $\geq 65$\%{} of $\Omega_{b}$(\ion{O}{6}). The large number of sightlines allows us for detection of strong \ion{O}{6}{} absorbers despite the low number-density of the lines, while also overcoming the problem of cosmic variance. While high-resolution, high S/N studies of QSO absorption lines are invaluable for IGM studies, we have shown here that low resolution studies also provide important and complementary results, as long as they cover a large number of sightlines.\\
Making the conservative assumption that the ionisation fraction, $\frac{OVI}{O}${}, can never be above 0.22 (for both the photoionised and the collisionally ionised cases), and adopting the Anders \& Grevesse solar abundance values, we derive for the mean metallicity of the gas probed in our search :\\
$\zeta (2.8 < z < 3.2) \geq 3.6 \times 10^{-4} h$\\
in good agreement with other studies. We take this fact, i.e. that the limits on number density, mass density {\bf and, to a lesser degree, the mean gas metallicity} we derive agree well with these studies, as further {\bf indication that our search strategy yields sensible results.}\\
Having obtained these observational data will allow us to confront them with simulations that model the intergalactic medium's metal enrichment, and serve as important constraint for some of the model parameters. We point out that a similar strategy can easily be applied to similar, even larger data-sets (as e.g. the BOSS survey project for SDSSIII will provide in the near future), which will enable without doubt much smaller (statistical) errors.\\

The SDSS is managed by the Astrophysical Research Consortium for the Participating Institutions. The Participating Institutions are the American Museum of Natural History, Astrophysical Institute Potsdam, University of Basel, University of Cambridge, Case Western Reserve University, University of Chicago, Drexel University, Fermilab, the Institute for Advanced Study, the Japan Participation Group, Johns Hopkins University, the Joint Institute for Nuclear Astrophysics, the Kavli Institute for Particle Astrophysics and Cosmology, the Korean Scientist Group, the Chinese Academy of Sciences (LAMOST), Los Alamos National Laboratory, the Max-Planck-Institute for Astronomy (MPIA), the Max-Planck-Institute for Astrophysics (MPA), New Mexico State University, Ohio State University, University of Pittsburgh, University of Portsmouth, Princeton University, the United States Naval Observatory, and the University of Washington. 

\bibliographystyle{aa}
\bibliography{sfrank}

\begin{thebibliography}{45}
\expandafter\ifx\csname natexlab\endcsname\relax\def\natexlab#1{#1}\fi

\bibitem[{{Aguirre} {et~al.}(2007){Aguirre}, {Dow-Hygelund}, {Schaye}, \&
  {Theuns}}]{aguirre2007}
{Aguirre}, A., {Dow-Hygelund}, C., {Schaye}, J., \& {Theuns}, T. 2007, ArXiv
  e-prints, 712

\bibitem[{{Aguirre} {et~al.}(2002){Aguirre}, {Schaye}, \&
  {Theuns}}]{aguirre2002}
{Aguirre}, A., {Schaye}, J., \& {Theuns}, T. 2002, \apj, 576, 1

\bibitem[{{Anders} \& {Grevesse}(1989)}]{anders1989}
{Anders}, E. \& {Grevesse}, N. 1989, \gca, 53, 197

\bibitem[{{Aracil} {et~al.}(2005){Aracil}, {Tripp}, \& {Bowen}}]{aracil2005}
{Aracil}, B., {Tripp}, T.~M., \& {Bowen}, D.~V. 2005, in IAU Colloq. 199:
  Probing Galaxies through Quasar Absorption Lines, ed. P.~{Williams}, C.-G.
  {Shu}, \& B.~{Menard}, 65--67

\bibitem[{{Bahcall} {et~al.}(1968){Bahcall}, {Greenstein}, \&
  {Sargent}}]{bahcall1968}
{Bahcall}, J.~N., {Greenstein}, J.~L., \& {Sargent}, W.~L.~W. 1968, \apj, 153,
  689

\bibitem[{{Bergeron} \& {Herbert-Fort}(2005)}]{bergeron2005}
{Bergeron}, J. \& {Herbert-Fort}, S. 2005, ArXiv Astrophysics e-prints

\bibitem[{{Burles} \& {Tytler}(1996)}]{burles1996}
{Burles}, S. \& {Tytler}, D. 1996, \apj, 460, 584

\bibitem[{{Carswell} {et~al.}(2002){Carswell}, {Schaye}, \&
  {Kim}}]{carswell2002}
{Carswell}, B., {Schaye}, J., \& {Kim}, T.-S. 2002, \apj, 578, 43

\bibitem[{{Cen} {et~al.}(1994){Cen}, {Miralda-Escud{\'e}}, {Ostriker}, \&
  {Rauch}}]{cen1994}
{Cen}, R., {Miralda-Escud{\'e}}, J., {Ostriker}, J.~P., \& {Rauch}, M. 1994,
  \apjl, 437, L9

\bibitem[{{Cowie} \& {Songaila}(1998)}]{cowie1998}
{Cowie}, L.~L. \& {Songaila}, A. 1998, \nat, 394, 44

\bibitem[{{Danforth} \& {Shull}(2007)}]{danforth2007}
{Danforth}, C.~W. \& {Shull}, J.~M. 2007, ArXiv e-prints, 709

\bibitem[{{Dav{\'e}} {et~al.}(1997){Dav{\'e}}, {Hernquist}, {Weinberg}, \&
  {Katz}}]{dave1997}
{Dav{\'e}}, R., {Hernquist}, L., {Weinberg}, D.~H., \& {Katz}, N. 1997, \apj,
  477, 21

\bibitem[{{Fox} {et~al.}(2008){Fox}, {Bergeron}, \& {Petitjean}}]{fox2008}
{Fox}, A.~J., {Bergeron}, J., \& {Petitjean}, P. 2008, \mnras, 388, 1557

\bibitem[{{Fox} {et~al.}(2007){Fox}, {Petitjean}, {Ledoux}, \&
  {Srianand}}]{fox2007}
{Fox}, A.~J., {Petitjean}, P., {Ledoux}, C., \& {Srianand}, R. 2007, \aap, 465,
  171

\bibitem[{{Fox} {et~al.}(2006){Fox}, {Savage}, \& {Wakker}}]{fox2006}
{Fox}, A.~J., {Savage}, B.~D., \& {Wakker}, B.~P. 2006, \apjs, 165, 229

\bibitem[{{Frank} {et~al.}(2010){Frank}, {Mathur}, \& {York}}]{frank2007}
{Frank}, S., {Mathur}, S., \& {York}, D.~G. 2010, ArXiv e-prints, 707

\bibitem[{{Haardt} \& {Madau}(1996)}]{haardt1996}
{Haardt}, F. \& {Madau}, P. 1996, \apj, 461, 20

\bibitem[{{Hellsten} {et~al.}(1998){Hellsten}, {Hernquist}, {Katz}, \&
  {Weinberg}}]{hellsten1998}
{Hellsten}, U., {Hernquist}, L., {Katz}, N., \& {Weinberg}, D.~H. 1998, \apj,
  499, 172

\bibitem[{{Hernquist} {et~al.}(1996){Hernquist}, {Katz}, {Weinberg}, \&
  {Miralda-Escud{\'e}}}]{hernquist1996}
{Hernquist}, L., {Katz}, N., {Weinberg}, D.~H., \& {Miralda-Escud{\'e}}, J.
  1996, \apjl, 457, L51+

\bibitem[{{Hu} {et~al.}(1995){Hu}, {Kim}, {Cowie}, {Songaila}, \&
  {Rauch}}]{hu1995}
{Hu}, E.~M., {Kim}, T., {Cowie}, L.~L., {Songaila}, A., \& {Rauch}, M. 1995,
  \aj, 110, 1526

\bibitem[{{Kim} {et~al.}(1997){Kim}, {Hu}, {Cowie}, \& {Songaila}}]{kim1997}
{Kim}, T., {Hu}, E.~M., {Cowie}, L.~L., \& {Songaila}, A. 1997, \aj, 114, 1

\bibitem[{{Kim} {et~al.}(2001){Kim}, {Cristiani}, \& {D'Odorico}}]{kim2001}
{Kim}, T.-S., {Cristiani}, S., \& {D'Odorico}, S. 2001, \aap, 373, 757

\bibitem[{{Kirkman} \& {Tytler}(1997)}]{kirkman1997}
{Kirkman}, D. \& {Tytler}, D. 1997, \apjl, 489, L123+

\bibitem[{{Lopez} {et~al.}(2007){Lopez}, {Ellison}, {D'Odorico}, \&
  {Kim}}]{lopez2007}
{Lopez}, S., {Ellison}, S., {D'Odorico}, S., \& {Kim}, T.~. 2007, ArXiv
  Astrophysics e-prints

\bibitem[{{Lu} {et~al.}(1996){Lu}, {Sargent}, {Womble}, \&
  {Takada-Hidai}}]{lu1996}
{Lu}, L., {Sargent}, W.~L.~W., {Womble}, D.~S., \& {Takada-Hidai}, M. 1996,
  \apj, 472, 509

\bibitem[{{Meyer} \& {York}(1987)}]{meyer1987}
{Meyer}, D.~M. \& {York}, D.~G. 1987, \apjl, 315, L5

\bibitem[{{Miralda-Escud{\'e}} {et~al.}(1996){Miralda-Escud{\'e}}, {Cen},
  {Ostriker}, \& {Rauch}}]{miraldaescude1996}
{Miralda-Escud{\'e}}, J., {Cen}, R., {Ostriker}, J.~P., \& {Rauch}, M. 1996,
  \apj, 471, 582

\bibitem[{{Nestor} {et~al.}(2008){Nestor}, {Hamann}, \& {Hidalgo}}]{nestor2008}
{Nestor}, D., {Hamann}, F., \& {Hidalgo}, P.~R. 2008, \mnras, 386, 2055

\bibitem[{{Petitjean} {et~al.}(1995){Petitjean}, {Mueket}, \&
  {Kates}}]{petitjean1995}
{Petitjean}, P., {Mueket}, J.~P., \& {Kates}, R.~E. 1995, \aap, 295, L9

\bibitem[{{Pieri} {et~al.}(2009){Pieri}, {Frank}, {Mathur}, {Weinberg}, {York},
  \& {Oppenheimer}}]{pieri2010}
{Pieri}, M.~M., {Frank}, S., {Mathur}, S., {et~al.} 2009, ArXiv e-prints, in
  press

\bibitem[{{Pieri} \& {Haehnelt}(2004)}]{pieri2004}
{Pieri}, M.~M. \& {Haehnelt}, M.~G. 2004, \mnras, 347, 985

\bibitem[{{Prochter} {et~al.}(2006){Prochter}, {Prochaska}, \&
  {Burles}}]{prochter2006}
{Prochter}, G.~E., {Prochaska}, J.~X., \& {Burles}, S.~M. 2006, \apj, 639, 766

\bibitem[{{Schaye} {et~al.}(2000){Schaye}, {Rauch}, {Sargent}, \&
  {Kim}}]{schaye2000}
{Schaye}, J., {Rauch}, M., {Sargent}, W.~L.~W., \& {Kim}, T.-S. 2000, \apjl,
  541, L1

\bibitem[{{Simcoe} {et~al.}(2002){Simcoe}, {Sargent}, \& {Rauch}}]{simcoe2002}
{Simcoe}, R.~A., {Sargent}, W.~L.~W., \& {Rauch}, M. 2002, \apj, 578, 737

\bibitem[{{Simcoe} {et~al.}(2004){Simcoe}, {Sargent}, \& {Rauch}}]{simcoe2004}
---. 2004, \apj, 606, 92

\bibitem[{{Simcoe} {et~al.}(2006){Simcoe}, {Sargent}, {Rauch}, \&
  {Becker}}]{simcoe2006}
{Simcoe}, R.~A., {Sargent}, W.~L.~W., {Rauch}, M., \& {Becker}, G. 2006, \apj,
  637, 648

\bibitem[{{Songaila}(1998)}]{songaila1998}
{Songaila}, A. 1998, \aj, 115, 2184

\bibitem[{{Steidel} \& {Sargent}(1992)}]{steidel1992}
{Steidel}, C.~C. \& {Sargent}, W.~L.~W. 1992, \apjs, 80, 1

\bibitem[{{Thom} \& {Chen}(2008{\natexlab{a}})}]{thom2008_I}
{Thom}, C. \& {Chen}, H.-W. 2008{\natexlab{a}}, ArXiv e-prints, 801

\bibitem[{{Thom} \& {Chen}(2008{\natexlab{b}})}]{thom2008_II}
---. 2008{\natexlab{b}}, ArXiv e-prints, 801

\bibitem[{{Tripp} {et~al.}(2008){Tripp}, {Sembach}, {Bowen}, {Savage},
  {Jenkins}, {Lehner}, \& {Richter}}]{tripp2008}
{Tripp}, T.~M., {Sembach}, K.~R., {Bowen}, D.~V., {et~al.} 2008, \apjs, 177, 39

\bibitem[{{Tytler} \& {Fan}(1994)}]{tytler1994}
{Tytler}, D. \& {Fan}, X.-M. 1994, in Bulletin of the American Astronomical
  Society, Vol.~26, Bulletin of the American Astronomical Society, 1424--+

\bibitem[{{Vanden Berk} {et~al.}(2008){Vanden Berk}, {Khare}, {York},
  {Richards}, {Lundgren}, {Alsayyad}, {Kulkarni}, {SubbaRao}, {Schneider},
  {Heckman}, {Anderson}, {Crotts}, {Frieman}, {Stoughton}, {Lauroesch}, {Hall},
  {Meiksin}, {Steffing}, \& {Vanlandingham}}]{vandenberk2008}
{Vanden Berk}, D., {Khare}, P., {York}, D.~G., {et~al.} 2008, \apj, 679, 239

\bibitem[{{Wild} {et~al.}(2008){Wild}, {Kauffmann}, {White}, {York}, {Lehnert},
  {Heckman}, {Hall}, {Khare}, {Lundgren}, {Schneider}, \& {Vanden
  Berk}}]{wild2008}
{Wild}, V., {Kauffmann}, G., {White}, S., {et~al.} 2008, ArXiv e-prints, 802

\bibitem[{{York}(2000)}]{york2000}
{York}, D.~G. e.~a. 2000, \aj, 120, 1579

\end{thebibliography}

\end{document}